\documentclass[pra,preprint]{revtex4}
\usepackage{amssymb}
\usepackage{bm,graphicx,amsmath}

\input{tcilatex}
\begin{document}

\title{Geometric phase of a moving three-level atom}
\author{S. Abdel-Khalek$^{1,2}$\email{sayedquantum@yahoo.co.uk}
 Y. S. El-Saman$^{1}$ and M. Abdel-Aty$^{1,3}$\email{mabdelaty@sci.uob.bh}}
\affiliation{$^{1}$Mathematics Department, Faculty of Science, Sohag University, Egypt \\
$^{2}$Math. Dept., Faculty of Science, Taif University, Saudi Arabia
\\
$^{3}$King Saud University, P.O. Box 2455, Riyadh, Saudi Arabia and
University of Bahrain, Bahrain}
\date{\today}

\begin{abstract}
In this paper, we investigate the geometric phase of the field interacting
with $\Xi $-type moving three-level atom. The results show that the atomic
motion and the field-mode structure play important roles in the evolution of
the system dynamics and geometric phase. We test this observation with
experimentally accessible parameters and some new aspects are obtained.
\end{abstract}

\maketitle


\section{Introduction}

In recent years much attention has been paid to the quantum phases such as
the Pancharatnam phase which was introduced in 1956 by Pancharatnam \cite%
{P56} in his studies of interference effects of polarized light waves. The
geometric phase which was realized, in 1984 \cite{B84}, is a generic feature
of quantum mechanics and depends on the chosen path in the space spanned by
all the likely quantum states for the system. Also, it has been shown that
such matrix element in quantum physics was proposed in path integral
approach to quantum mechanics \cite{saw22}. This approach studies the
multitude of quantum trajectories which connect two points in the Hilbert
space: $\psi (q^{j}(0))$ and $\psi (q^{j}(t))$, where $q^{j}$ are
generalized quantum coordinates. The definition of phase change for partial
cycles was obtained by Jordan \cite{J88} and ideas of Pancharatnam were used
\cite{SB88,BM87} to show that for the appearance of \textrm{Pancharatnam's}
phase the system needs to be neither unitary nor cyclic \cite{WB90,WL88},
and may interpreted by quantum measurements.\textrm{\ In this regard, one
should noticed that the geometric phase is different from other phase
effects in cavity quantum electrodynamics. For example tangent and cotangent
states \cite{slo89} are connected with the phase of atomic population in two
level system and the same phase effects are proposed in three-level system
\cite{ors92} .}

Presently the models of quantum computation in which a state is an operator
of density matrix are developed \cite{T02}. It is shown \cite{EEHI00} that
the geometric phase shift can be used for generation fault-tolerance phase
shift gates in quantum computation. Many generalizations have been proposed
to the original definition \cite{VM00,A03,AAO00,Law99}. The quantum phase,
including the total phase as well as its dynamical and geometric parts, of
Pancharatnam type are derived for a general spin system in a time-dependent
magnetic field based on the quantum invariant theory \textrm{\cite{wag95}}.
Another approach that provides a unified way to discuss geometric phases in
both photon (massless) and other massive particle systems was developed by
Lu in Ref. \textrm{\cite{lu99}}. Also, an expression for the Pancharatnam
phase for the entangled state of a two-1evel atom interacting with a single
mode in an ideal cavity with the atom undergoing a two-photon transition was
studied \textrm{\cite{Law99}}. To bring the two-photon processes closer to
the experimental realization, the effect of the dynamic Stark shift in the
evolution of the Pancharatnam phase has been presented \cite{AAO00}. More
recently, a method for analyzing the geometric phase for $N$ two-level
system of superconducting charge qubits interacting with a microwave field
is proposed \cite{aty09} and through a simple but universal system
(two-level atom) a possibility to control the Pancharatnam phase of a
quantum system on a much more sensitive scale than the population dynamics
has been reported \cite{bou09}. In Ref. \cite{P03} experiments are proposed
for the observation of the nonlinearity\ of the Pancharatnam phase with a
Michelson interferometer.

In this paper we extend these investigations to study the dynamics of a
moving three-level atom interacting with a coherent field. An exact solution
of a three-level atom in interaction with cavity field has been obtained
\cite{chu82} and developed for arbitrary configuration of the levels \cite%
{li87}. We investigate the effect of different parameters of the system on
the geometric phase. This paper is arranged as follows: In Sec. 2, we
introduce the model and its solution by using the unitary transformations
method. In Sec. 3, we investigate the geometric phase and the dynamical
properties for different regimes. Numerical results for the geometric phase
are discussed in Sec. 4. Finally conclusions are presented.

\section{The system}

In this section, we discuss the model of a moving three-level atom with
energy levels denoted by $|1\rangle ,|2\rangle $ and $|3\rangle ,$ where $%
|3\rangle $ is the ground level, $|2\rangle $ is the middle level and $%
|1\rangle $ is the upper level. The interaction Hamiltonian of the system in
the rotating-wave approximation can written as \cite{yoo85}
\begin{equation}
H_{I}=g\lambda (t)\left( a\sigma _{23}e^{i\Delta t}+a\sigma _{12}e^{-i\Delta
t}+h.c.\frac{{}}{{}}\right) ,
\end{equation}%
where $a(a^{\dagger })$ are the annihilation (creation) operator of the
field mode, $\sigma _{ij}=|i\rangle \langle j|$ are the atomic operators, $%
\Delta $ is the detuning of the field mode from the atomic middle level 2 $%
(\Delta =\omega -(\omega _{2}-\omega _{1})=\omega -(\omega _{2}-\omega _{3}))
$ and $g$ is the atom--field coupling constant$.$ We deal with the
one-dimensional case of atomic motion along the cavity axis and denote by $%
\lambda (t)$ the shape function of the cavity field mode \cite%
{sar74,li87,ena08,sch89}. A realization of particular interest in which the
atomic center-of-mass motion is classical can be written as $\lambda
(t)=\sin (p\pi \upsilon t/L),$ where $\upsilon $ denotes the atomic motion
velocity, $p$ stands for the number of half wavelengths of the mode in the
cavity and $L$ is the cavity length in $z-$direction. The classical
consideration of the atomic center-of-mass motion is very well obeyed under
the ordinary experimental conditions \cite{eng94}. However under extreme
circumstances the quantum nature of the center-of-mass motion may become
important \cite{eng91,har91}. In Ref. \cite{eng94} it has been mentioned
that in the time dependent evolution the integral of $\lambda (t)$\ must be
approximate with $\overline{\lambda }t$, where $\overline{\lambda }$
describes the mean value of the interaction constant between the atom and
cavity mode.

In this paper we assume that, the initial state is given by $|\psi
(0)\rangle $ $=|\psi _{A}(0)\rangle \otimes |\psi _{F}(0)\rangle ,$ where $%
|\psi _{A}(0)\rangle $ is the initial state of the atom and $|\psi
_{F}(0)\rangle $ is the initial state of the field. The combined atom-field
system can be written as \cite{rei09}

\begin{equation}
|\psi (0)\rangle =\sum_{l=0,1}\cos \left( \theta +\frac{l\pi }{2}\right)
\left( \sum_{n=-l}^{\infty }\frac{q_{n+l}(1+r(-1)^{n+l})}{\sqrt{B}}|n+l\text{
}\rangle _{F}\right) |3-l\text{ }\rangle _{A},  \label{ini}
\end{equation}%
where $B=(1+r^{2}+re^{-2n})$, $r$ is an arbitrary constant. In this case, $%
r=0,1,$ and $-1,$ corresponds to coherent state, even coherent and odd
coherent state, respectively. While $q_{n}$ is the number-state expansion
coefficient, $q_{n}=\langle n|\psi _{F}(0)\rangle ,$ for coherent state $%
q_{n}=e^{-\bar{n}/2}\alpha ^{n}/\sqrt{n!},$ and $\bar{n}=|\alpha |^{2}$ is
the average photon number of the field.

\section{Geometric phase}

For the quantum system evolving from an initial wavefunction to a final
wavefunction, if the final wavefunction cannot be obtained from the initial
wavefunction by a multiplication with a complex number, the initial and
final states are distinct and the evolution is noncyclic. Suppose state $%
\left\vert \psi (0)\right\rangle $ evolves to a state $\left\vert \psi
(t)\right\rangle $ after a certain time $t$. If the scalar product \cite%
{aha00}\textrm{\ }
\begin{equation}
\left\langle \psi (0)\right\vert \exp \left[ \frac{i}{\hbar }%
\int_{0}^{t}H(t^{\prime })dt^{\prime }\right] \left\vert \psi
(0)\right\rangle ,
\end{equation}%
can be written as $\Gamma \exp (i\gamma ),$ where $\Gamma $ is a real
number, then the noncyclic phase due to the evolution from $\left\vert \psi
(0)\right\rangle $ to $\left\vert \psi (t)\right\rangle $ is the angle $%
\gamma .$ This noncyclic phase generalizes the cyclic geometric phase since
the latter can be regarded as a special case of the former in which $\Gamma
=1$. Determination of the phase between the two states for such an evolution
is nontrivial. Pancharatnam prescribed the phase acquired during an
arbitrary evolution of a wavefunction from the vector $\left\vert \psi
(0)\right\rangle $ to $\left\vert \psi (t)\right\rangle $ as
\begin{equation}
\phi _{g}=\mathrm{arg}\left\langle \psi (0)\right\vert \left. \psi
(t)\right\rangle .
\end{equation}%
Subtracting the dynamical phase from the Pancharatnam phase, we obtain the
geometric phase. Here, for the time-dependent interaction and considering
the resonant case, an exact expression of the geometric phase can be
obtained as
\begin{equation}
\phi _{g}(t)=-\sin ^{-1}\left\{ \frac{y(t)}{\sqrt{x^{2}(t)+y^{2}(t)}}%
\right\} ,
\end{equation}%
where
\begin{eqnarray}
x(t) &=&\sum_{n=0}^{\infty }\frac{q_{n}^{2}\cos ^{2}\theta }{B(2n+3)}%
[1+r(-1)^{n}]^{2}[n+2+(n+1)\cos (\lambda _{1}(t)\sqrt{2n+3})]  \notag \\
&&+\sum_{n=0}^{\infty }\frac{q_{n+1}^{2}\sin ^{2}\theta }{B}%
[1-r(-1)^{n}]^{2}\cos \left[ \lambda _{1}(t)\sqrt{2n+3}\right] , \\
y(t) &=&\sum_{n=0}^{\infty }\frac{q_{n}q_{n+1}\sin 2\theta }{B}\sqrt{\frac{%
n+1}{2n+3}}[1-r^{2}]\sin \left[ \lambda _{1}(t)\sqrt{2n+3}\right] , \\
\lambda _{1}(t) &=&\frac{L}{p\pi \upsilon }(1-\cos \left( p\pi \upsilon
t/L\right) ).\   \notag
\end{eqnarray}%
More specifically, if we consider the atomic motion velocity as $\upsilon
=gL/\pi $, then $\lambda _{1}(t)=\frac{1}{pg}(1-\cos \left( pgt\right) ),\ $%
while $\ \lambda _{1}(t)=gt,$\ if the atomic motion is neglected.

For the off-resonant case, the numerical results will be used. In the
time-independent case, characterized by the transformation matrix $\hat{U}%
(t),$\ the geometric phase is defined through $|\psi (t)\rangle =\hat{U}%
(t)|\psi (0)\rangle $ and $\hat{U}(t)$\ is given by
\begin{equation}
\hat{U}(t)=\sum_{j=1}^{3}\exp \left( -iE_{j}t\right) |\varphi _{j}\left(
t\right) \rangle \langle \varphi _{j}\left( t\right) |,  \label{4}
\end{equation}%
where $|\varphi _{j}\left( t\right) \rangle $ and $E_{j\text{ }}\left(
t\right) $ are eigenvectors and corresponding eigenvalues of the Hamiltonian
$H_{I}.$

\subsection{Numerical results}

It is of rather more use to exhibit the numerical results explicitly for
particular initial conditions of relevance to the experiments. With this in
mind we will assume that the initial state is prepared according to Eq. (\ref%
{ini}) to be a particular coherent state, even coherent or odd coherent
state, with the atom prepared in a superposition state. In the experiments
done so far, it is possible to probe directly which electronic state the
atom occupies. It was reported that \cite{bru96} the cavity can have a
photon storage time of $T=1$ ms (corresponding to $Q=3\times 10^{8}$ ). The
radiative time of the Rydberg atoms with the principle quantum numbers $49$,
$50$ and $51$ is about $2\times 10^{-4}s$. In order to realize such a scheme
in laboratory experiment within a microwave region, we may consider slow $Rb$
atoms in higher Rydberg states which have life time of the order of few
milliseconds \cite{dav94}. The interaction times of atom with the cavity
modes come out to be of the order of few tens of microseconds which is far
less than the cavity life time. The high-Q cavities of life time of the
order of millisecond are being used in experiments \cite{rau} which opened
interesting perspectives for quantum information manipulation and
fundamental tests of quantum theory.

\begin{figure}[tbp]
\centering\includegraphics[width=1\textwidth]{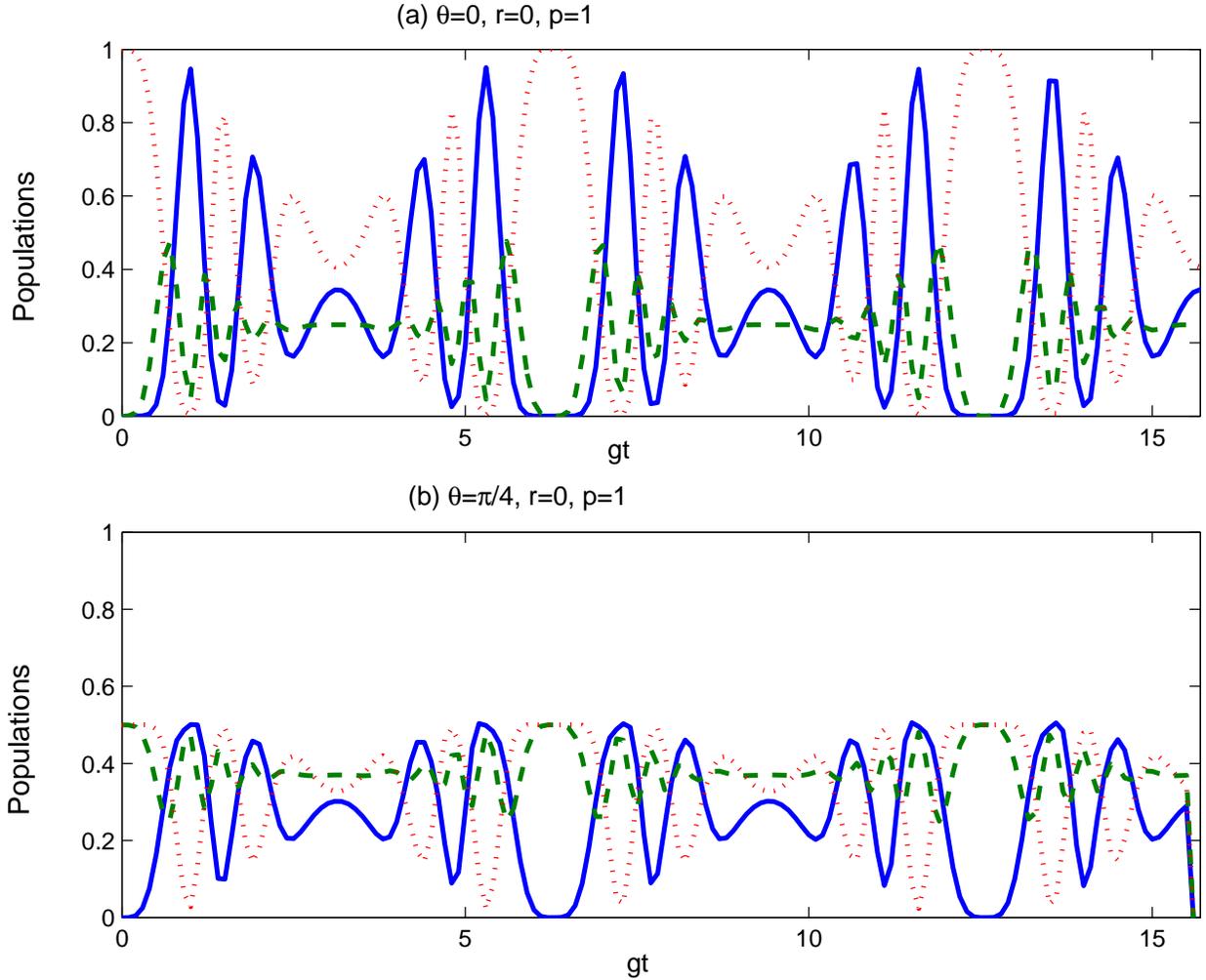}
\caption{Time evolution of the populations $\protect\rho _{11}$ (solid
line), $\protect\rho _{22}$ (dashed line), $\protect\rho _{33}$ (dotted
line) for $\protect\alpha =5$, $\Delta =0$, $r=0,p=1$ and different values
of $\protect\theta ,$ where (a) $\protect\theta =0$ and (b) $\protect\theta =%
\protect\pi /4$. }
\end{figure}

It is clear that the geometric phase has zero value when $\theta =0$, this
means physically that there is no phase in photon transition that make the
photon transition is in a straight line and populations have values of
oscillations between $0$ and $1.$ As a result of the numerical calculations,
the oscillations will dephase and next will collapse after some times $t$ .
In figure 1, we present a plot in which a comparison between the general
behavior of atomic dynamic when $\theta =0$ and $\theta =\pi /4$ is
presented. It is shown that for $\theta =0,$ the oscillations of the
probability amplitudes are repeated periodically, the oscillations have
peaks and bottoms where populations are vanished every time periodically at $%
t=\frac{2\pi }{p},p=1$. On the other hand, when $\theta =\pi /4,$ we observe
that the amplitude of the oscillations is decreased and the maximum value of
the populations does not exceed $0.5$.\bigskip

\begin{figure}[tbp]
\centering\includegraphics[width=1\textwidth]{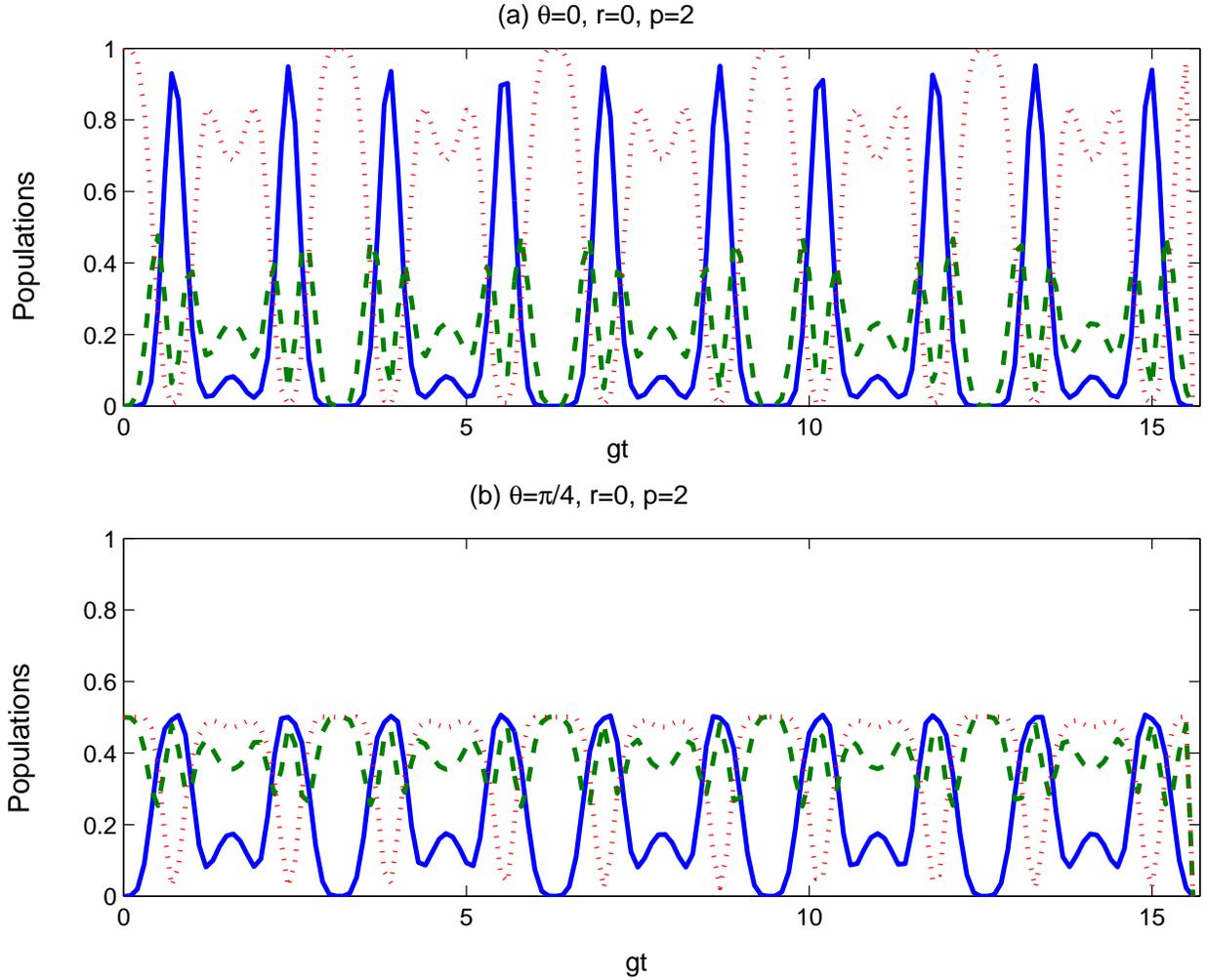}
\caption{The same as figure 1 but $p=2.$}
\end{figure}

In figure 2 we compare the populations when $\theta =0$ and with that when $%
\theta =\pi /4$ in the same period of time (see figure 1). For $\theta =0,$
we still have the periodicity of the dynamics but in this case the vanishing
time of the populations is given by $t=\frac{2\pi }{p},\left( p=2\right) $.
Similar to the previous case for $\theta =\pi /4,$ the amplitude of the
oscillations is decreased and the maximum value of the populations does not
exceed $0.6$.

In Fig. 3a we have plotted the geometric phase $\phi _{t}$ as a function of
the scaled time, $gt,$ where $r=0$ and $p=1.$ When $\theta =0$ we find that
the geometric phase is vanished but for $\theta =\pi /4$ the geometric phase
shows similar behavior to the periodic collapse-revival phenomenon of Rabi
oscillation but with a period of ($t=\frac{2\pi }{p},p=1$). In Fig. 3b we
have plotted the geometric phase, $\phi _{t\text{ }},$ as a function of the
scaled time, $gt,$ \ where $r=0$ and $p=2.$ In this case, it is shown that
the geometric phase has the same behavior of Fig. 3a but with a period of ($%
t=\frac{2\pi }{p},p=2$). In the off-resonant case, the geometric phase shows
oscillatory behavior only during the first stage of the interaction (say $%
0\leq \lambda t\leq 10),$ followed by zero phase.

\begin{figure}[tbp]
\centering\includegraphics[width=1\textwidth]{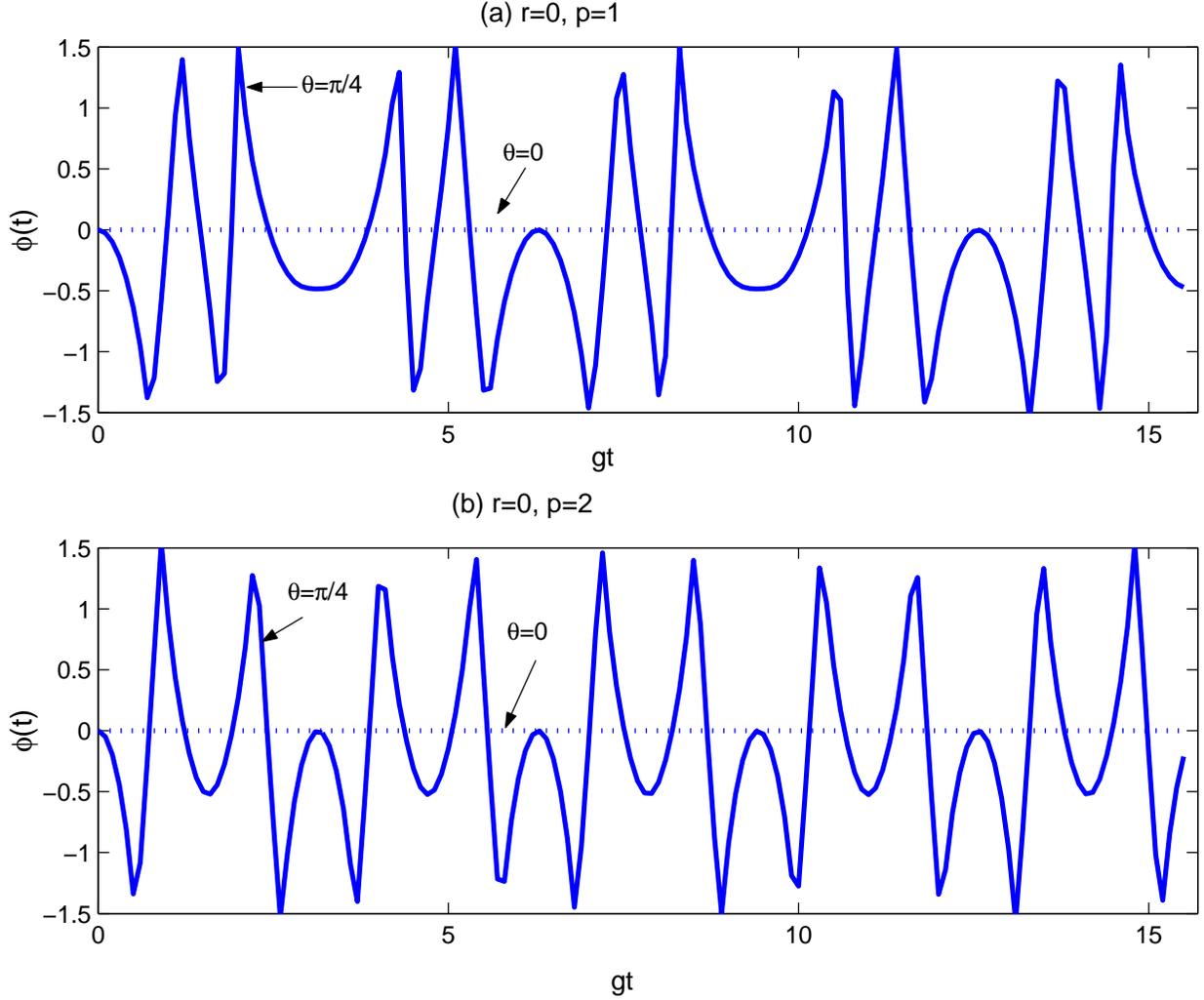}
\caption{The geometric phase, $\protect\phi \left( t\right) ,$ as a function
of the scaled time $gt$. The other parameters are $\protect\alpha =5$, $%
\Delta =0,$ $r=0$ and different values of $p$ and $\protect\theta $ where
(a) $p=1$ and (b) $p=2$ with $\protect\theta =0$ (dotted curve) and $\protect%
\theta =\protect\pi /4$ (solid curve)}
\end{figure}
\begin{figure}[tbp]
\centering\includegraphics[width=1\textwidth]{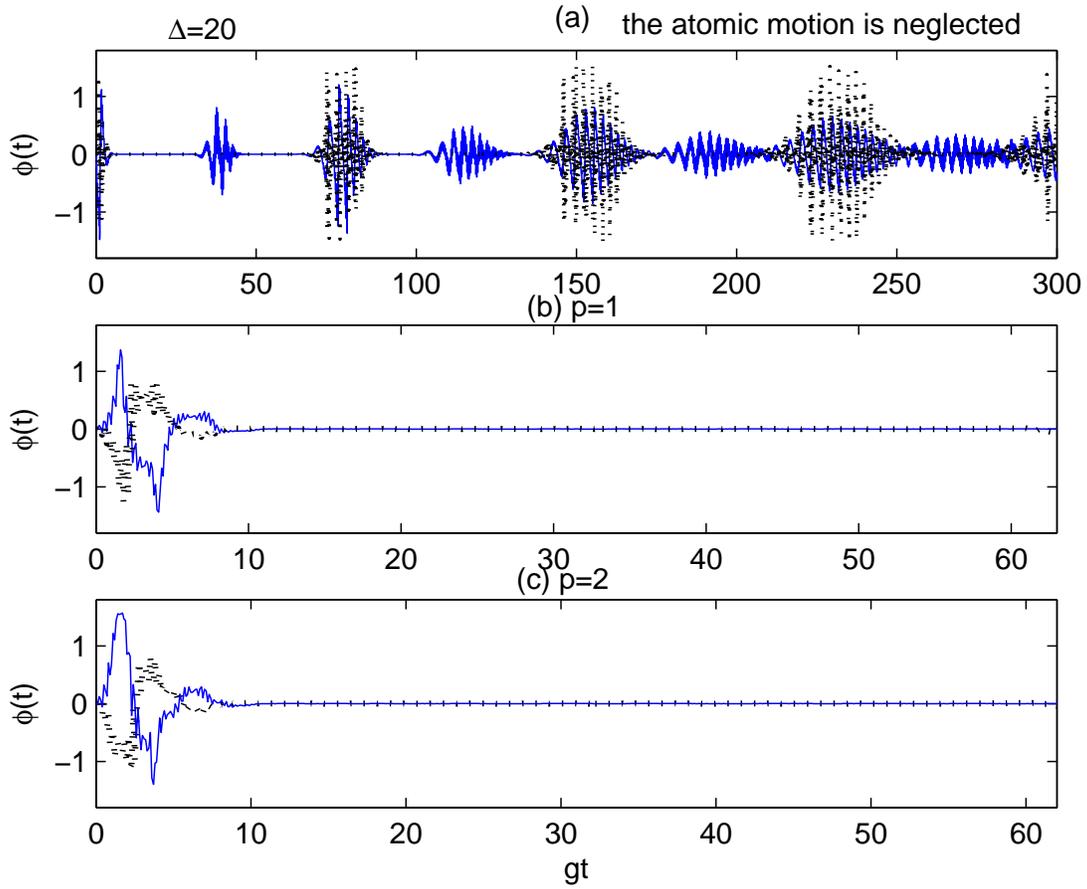}
\caption{Time evolution of the geometric phase $\protect\phi \left( t\right)
$ for $\protect\alpha =5$, $\Delta =20$, $\protect\theta =\protect\pi /4$
where (a) the atomic motion is neglected, (b) $p=1$, (c) $p=2$ with $r=1$
(dotted line) and $r=0$ (solid line).}
\end{figure}

In Fig. 4, we have plotted the geometric phase when we consider the
influence of the detuning, $\Delta =20,$ and for different values of the
parameter $r$ , where $r=0$ (solid line) and $r=1$ (dotted line ). In Fig.
4a the atomic motion is neglected $p=0.$ It is important to note that the
geometric phase has a collapse-revival but the amplitude of the oscillations
as well as the revival time become smaller and we have more revivals at the
same period of time time and the two values of revival geometric phase which
are corresponding to $r=0$ and $r=1$ repeating with the time development. In
Fig. 4b, we consider $p=1$, keeping the same value of the parameter $r=1$ as
in Fig. 4a. It is shown that the geometric phase has oscillations only at
the initial stage of the interaction time. \ As time goes on ($gt>8$), a
disappearance of the geometric phase is observed and there is no more
dependence on the interaction time.

\begin{figure}[tbp]
\centering\includegraphics[width=1\textwidth]{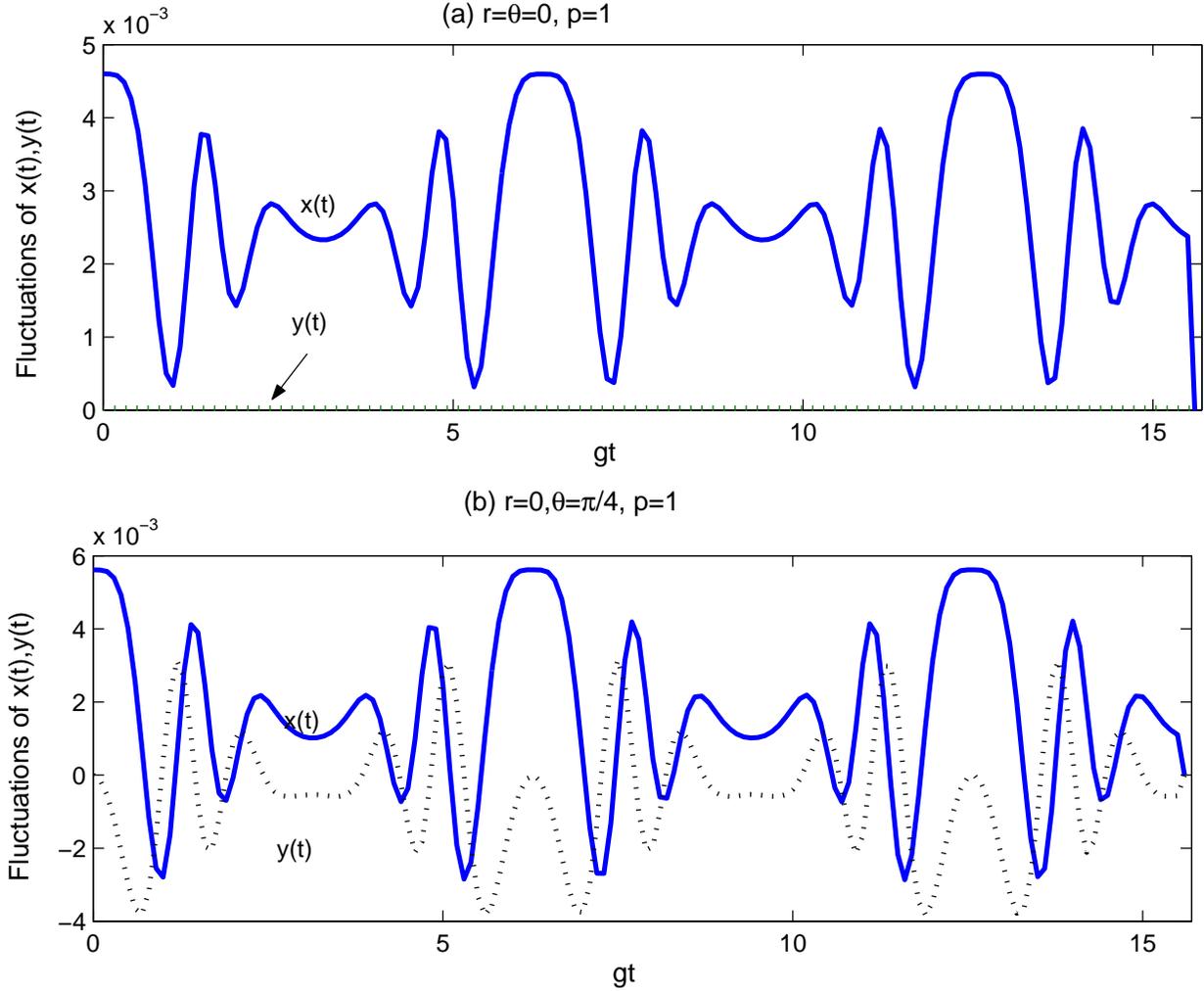}
\caption{Time evolution of $x(t)$ (soled curve) and $y(t)$ (dotted curve),
where $\langle \protect\psi (0)|\protect\psi (t)\rangle =x(t)+iy(t).$ The
parameters are $\protect\alpha =5$, $\Delta =0,$ $r=0,p=1$ and different
values $\protect\theta $ where (a) $\protect\theta =0$ and (b) $\protect%
\theta =\protect\pi /4.$}
\end{figure}

It is rather interesting to mention to the fact that: in the Feynman
approach to quantum mechanics it is important not only the modulus and
argument of complex number, but the quantum fluctuations of this variables
too. In such a representation, the deviation of quantum trajectories from
the classical trajectories are described by the quantum fluctuations \cite%
{elu09}. Therefore we have further investigated the fluctuations of the real
and imaginary parts of $\left\langle \psi (0)\right\vert \left. \psi
(t)\right\rangle =x(t)+iy(t)$ for two different values of $\theta ,$ where $%
\theta =0$ or $\theta =\pi /4$ (see Fig. 5). Fig. 5a gives a clear physical
picture in this case for the disappearance of the geometric phase when $%
\theta =0,$ since $x(t)$ shows oscillations only while $y(t)=0$ everywhere.
However, if the influence of the geometric phase on the parameter $\theta $
is to be included, e.g., $\theta =\pi /4$ (see Fig. 5b), oscillations are
clearly observed for both $x(t)$ and $y(t)$ and hence the geometric phase
exists in this case (see Fig. 3).

\section{conclusion}

In this paper we have investigated the quantum dynamics and geometric phase
of the interaction between a moving three-level atom and a coherent field.
We have used the unitary transformation method to obtain an exact expression
of the geometric phase and numerical treatment has been used for the
off-resonant case. The results point to a number of interesting features,
which arise from the variation of the parameters of the system, namely, the
atomic motion, and atomic superposition parameter. This result comes down to
say that every parameter has an effect on the geometric phase. If the atom
starts from an upper state, we find that there is no value of the geometric
phase which means physically that the photon transition will occurs without
any phase. Meanwhile, if the atom starts from a superposition case, the
geometric phase will oscillates periodically depending on the value of
atomic motion parameter $p.$

\bigskip

{\large Acknowledgement}

The authors would like to thank the referees for their objective comments
that improved the text in many points.

\end{document}